\documentclass[numreferences]{kluwer}    
\usepackage[]{graphicx,klucite}

\begin{document}                                                                                   
\begin{article}
\begin{opening}         
\title{220 GHZ ZENITH ATMOSPHERIC TRANSPARENCY AT IAO, HANLE 
\thanks{Hanle :  Latitude $32^{\circ}46^{\prime}46^{\prime\prime}$ N; 
Longitude $78^{\circ}57^\prime 51^{\prime\prime}$ E; Altitude 4500~m}} 

\author{P.G. Ananthasubramanian$^1$, Satoshi Yamamoto$^2$,\\ and Tushar P.
Prabhu$^3$}
  
\institute{\em $^1$Raman Research Institute, Bangalore 560 080, India\\
$^2$Dept. of Physics, University of Tokyo, Bunkyo-ku, Tokyo 113, Japan\\
$^3$Indian Institute of Astrophysics, Bangalore 560 034, India.}
\runningauthor{P.G. Ananthasubramanian et al.}
\runningtitle{220 GHz Atmospheric Transparency at IAO, Hanle}
\end{opening}

\section*{Abstract}
\begin{minipage}{24pc}
\begin{quote}
We present 220~GHz (1.36~mm) measurements of zenith optical depth 
obtained to characterise the Indian Astronomical Observatory, Hanle 
(Ladakh, India) during the period from late December 1999 to early May 
2000 and early October 2000 to September 2001. The data were sampled at 
an interval of 10 minutes.  We describe the automated 220~GHz tipping 
radiometer used, its basic principle, operation, data acquisition method
and data reduction scheme in detail. The 220~GHz opacity is found to be 
less than 0.06 for a significant fraction (40\%) of the time during the 
winter months, indicating that Hanle is one of the good observing sites 
for submillimeter-wave astronomy.  We make a preliminary correlation 
with the precipitable water vapour derived from surface relative 
humidity and air temperature measurements made during the same period 
with a weather station installed at the site. We also compare the Hanle 
site with other high-altitude sites like Mauna Kea and Atacama desert. 

{\bf keywords:} atmospheric opacity, precipitable water vapour ($pwv$), 
surface water vapour pressure($P_o$), super-heterodyne receiver, sub-mm 
astronomy and mm-wave astronomy 
\end{quote}
\end{minipage}           

\section*{\underbar{Introduction}}

Until recently, the only well-established high-altitude astronomical 
site in the northern hemisphere was Mauna Kea, Hawaii (altitude 4100~m).
Mt Evans (4300~m) in Colorado is currently being developed by the 
University of Denver. The Indian Astronomical Observatory (IAO) was set 
up recently in the high-altitude cold desert of Changthang 
(``highlands'') Ladakh \cite{HIROT}  with a 2-m aperture optical-infrared
telescope \cite{Anupama, Cowsik-etal}. IAO is atop Mt.\ Saraswati 
(4500~m), Digpa-ratsa Ri, a range of hillocks surrounded by vast plains 
(Nilamkhul Plain at 4300~m). 

The electromagnetic spectrum in the sub-mm region is not so well 
explored for astronomy as yet. Apart from the JCMT and CSO at Mauna Kea 
in the northern hemisphere, there are only a few other facilities setup 
for observations in this wavelength band, e.g.  Gornergrat, in 
Switzerland (altitude 3167~m) and Mt. Fuji in Japan (3776~m). A high 
altitude site (4000--5000~m) has been identified in Chile for upcoming 
projects such as the Atacama Large Millimeter Array (ALMA).  This is a 
well characterised site in the southern hemisphere apart from the South 
Pole, Antarctica (2835~m). There are other high-altitude sites proposed
for mm-wave astronomy, such as Cerro La Negra (4600~m) in Mexico. The 
need for a high-altitude site arises primarily because atmospheric 
constituents like water vapour, oxygen and a few other molecules absorb 
a substantial portion of the incoming radiation in the sub-mm spectral 
region and such absorption reduces as one goes to higher altitudes. The 
atmosphere not only reduces the signal strength but also contributes to 
an increase in system temperature and thus doubly degrades the system 
sensitivity. Of the various atmospheric constituents, oxygen content is 
fairly constant both diurnally and seasonally for a given location, and 
depends on its latitude and altitude. The water vapour content, on the 
other hand, varies both diurnally and seasonally. There are some 
semitransparent windows between the emission/absorption bands, which 
lie in the troughs formed between the wings of the bands. The 
trasmission in these wings improves with altitude.

There are several emission lines and bands of astrophysical interest in
the available windows: 346~GHz (CO $J$ = 3--2), 460~GHZ (CO $J$=4--3), 
492~GHz (C~I), 660~GHz ($^{13}$CO $J$=6--5), 692~GHz (CO $J$=6--5), 
806~GHz (CO $J$=7--6) and 880~GHz ($^{13}$CO $J$=8--7). Ideally, it will
be important to monitor any prospective site for its atmospheric 
transmission quality in these windows before a large investment is made 
for a major facility. Since sub-mm instrumentation is very expensive and
requires considerable logistic support, it has been a general practice 
to monitor the atmospheric transmission at lower frequencies such as 
183, 220 or 225~GHz in the mm-wave region.  The transparencies in the 
higher frequency windows can be scaled using suitable models of earth's 
atmosphere. Some of these measurements are discussed by 
\cite{Radford-Chamberlin, Matsushita-etal, Hirota-etal, Matsuo-etal, 
Chamberlin-etal, Holdaway-etal, Sekimoto-etal}.  

The identification and development of the Hanle site amidst a vast cold
desert landscape, where the ambient temperatures go down to $-25^\circ$
C during winter nights,  opens up possibilities for setting up of future
large observing facilities in the sub-mm and lower wavelength bands in 
the northern hemisphere.  The atmospheric transparency at the sub-mm
frequencies are expected to be very good judging from the surface 
relative humidity and air temperature data that are being recorded at 
Hanle since the summer of 1996.  However, it is indispensable to measure
the atmospheric opacity directly at the mm-wave frequency in order to
characterize the actual observing conditions at the site.  Such opacity
measurements have been reported for other high altitude sites, and we 
can compare the results of Hanle with these quantitatively. With this in
mind, a team of Japanese astrophysicists visited the site and measured
the atmospheric opacity at 220~GHz in November 1996 . At that time, the 
220~GHz opacity was around 0.05, indicating good transparency even at 
sub-mm frequencies.  Encouraged by this experiment, it was decided to 
operate the 220~GHz radiometer continuously at the site for a 2-year 
period. In this paper, we describe the first long-term results of the 
220~GHz opacity measurements at IAO, Hanle.

\section*{\underbar{Description of the 220~GHz system}}

\begin{figure} 
\vspace{1pc}
\centerline{\includegraphics[width=24pc]{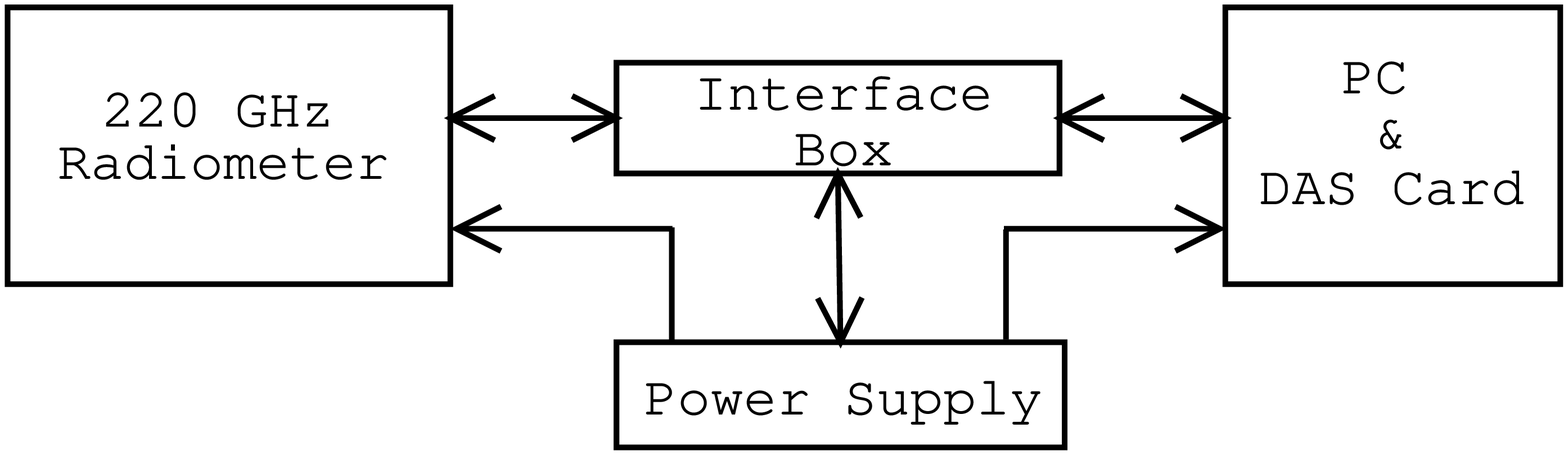}} 
\vspace{1pc}
\caption[]{A block diagram of the radiometer system showing all the
four different units. Only the 220~GHz radiometer is installed outdoors 
on a pedestal at about 3~m above the ground. The other three units are 
housed inside the seeing monitor building located near the pedestal
(see Fig.~3).} 
\label{rmblk}
\end{figure}

The radiometer is essentially the same unit as used at Mt.~Fuji during
1994--95 \cite{Sekimoto-etal}. The system consists of four distinct 
units: (i) a 220~GHz receiver with a stepper-motor-driven off-axis 
paraboloid to scan in elevation, (ii) a power supply, (iii) an interface
unit and (iv) a personal computer running on DOS with a hard disk for 
data storage (Fig.~1). The PC communicates with both these units through
an add-on card. The unit (i) is meant for outdoor use with a metal 
radome top of semi-cylindrical shape.  The radome has a rectangular 
window covered with a layer of low-loss woven teflon membrane which is 
transparent to mm-waves. This ensures continuous safe operation of the 
electronics in the outdoor environment. The inner surface of the radome 
is packed with 50~mm thick thermal insulating material to provide some 
isolation from the outside sub-zero ambient temperatures. Heating of the
radome was not considered necessary since snowfall is rare and 
negligible at the site. The small amount of wet snow in autumn and
spring melts away in the subsequent sunlight and the dry snow in deep 
winter is blown away by the wind. The box containing the intermediate 
frequency (IF) electronics and the detector housed inside the radome 
unit is heated and maintained at $7^{\circ}$ C ($T_{heater}$). The 
program maintains the IF box temperature by on-off control during the 
time between the 10 minute sky scans, should the temperature fall below 
this (Fig.~2).

\begin{figure} 
\vspace{1pc}
\centerline{\includegraphics[width=24pc]{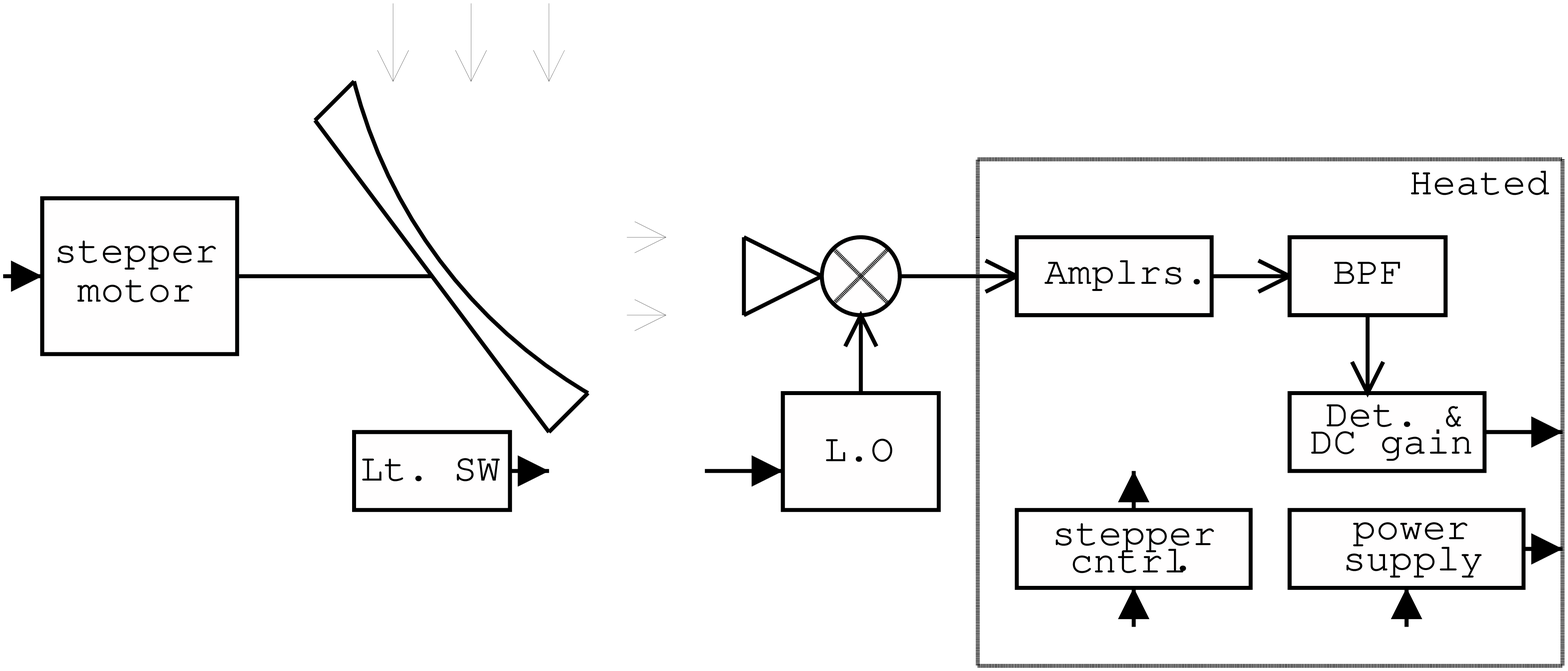}} 
\vspace{1pc}
\caption[]{A schematic diagram showing some of the mm-wave and 
other electronics housed in the radome. The mirror scans the sky in 
elevation through a layer of protective membrane fixed to the 
rectangular window on the radome.} 
\label{rmsch}
\end{figure}

The off-axis parabolic mirror of focal length 150~mm, with a projected 
aperture of 80~mm, is mounted at 45$^{\circ}$ to the rotation axis, and 
the 220~GHz prime-focus receiver is fixed on the rotation axis. The 
rotation axis is horizontal.  The reflected and focussed beam is normal 
to the incoming radiation and lies along the rotation axis of the mirror
and stepper motor as the mirror scans the sky from one horizon through 
the zenith to nearly the opposite horizon. With this simple arrangement,
the system can look at the sky at all elevations or zenith angles ($z$) 
at a fixed azimuth position, with just one axis rotation and no 
blockage. The effective beam size on the sky is about 1$^{\circ}$. 

The 220~GHz receiver is a super-heterodyne receiver consisting of a 
corrugated conical horn, a sub-harmonically pumped Schottky diode mixer 
and a Gunn diode local oscillator (LO) working at 110~GHz. The 
down-converted signal centred at 1450~MHz is amplified by a low-noise 
amplifier and a post-amplifier, and passed through a band-pass filter 
(BPF) to limit the bandwidth to 500~MHz.  The band-limited signal is 
square-law detected and amplified for digitising. Digitising is done by 
a commercial PC add-on analog-to-digital converter and digital 
input/output card which is also used for controlling the stepper motor 
movement. There is no encoder to identify the beam position on the sky. 
A sky scan starts from a home position (zenith angle about$-90^{\circ}$)
identified with a limit-switch status. A typical scan starts from the 
home position and ends at the home position, passing through the zenith 
position.  The program also monitors system power supply voltages and 
temperatures inside the radome and on the IF box (Fig.~2).

The mirror with a beam size about $1^{\circ}$ on sky is driven by a 
stepper motor under PC control. It is programmed to look at every 
$0.72^{\circ}$ in zenith angle on the sky for 90~ms, covering a zenith 
angle range of nearly $-90^{\circ}$ to $+70^{\circ}$.  The sky coverage 
through the radome window is limited to about $90^{\circ}$, i.e., from 
about $-75^{\circ}$ to $+15^{\circ}$ in zenith angle. A forward scan and
a reverse scan are taken in about 50~s. Between $+22^{\circ}$ and
$+60^\circ$ zenith angle position is a reference load, a blackbody at 
room temperature ($T_{ref}$). The averaged signal corresponding to these
mirror positions is the reference for temperature calibration of the 
signal received from the sky emission. Data are recorded with time stamp
along with the monitored parameters as well as the fitted opacity 
values. A separate file is also generated, one for a day, wherein the 
time stamp, raw data file name, fitted opacity and some monitored system
parameter values are recorded for every pair of scans. The program is 
configured to take about 144 scans in 24 hours, i.e., a pair of scans 
every 10 minutes.

\begin{figure} 
\vspace{-21pc}
\centerline{\includegraphics[width=50pc]{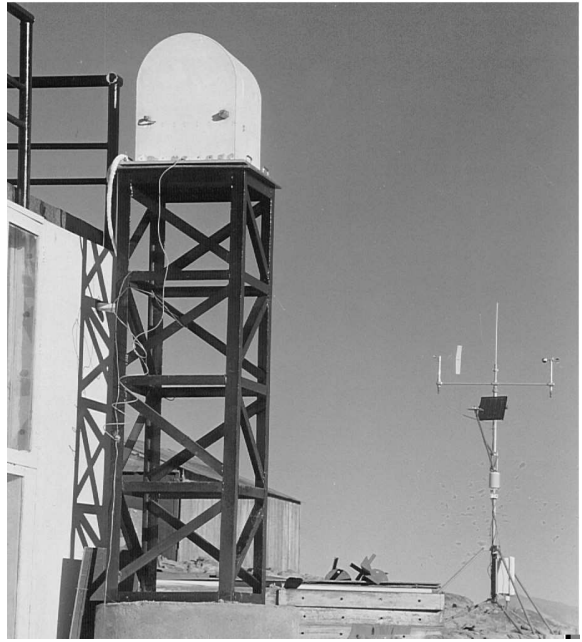}} 
\vspace{-25pc}
\caption[]{A view of the 220~GHz radiometer on the pedestal by the side 
of the seeing monitor building. Also seen is the weather station.}
\label{rmpic}
\end{figure}

The instrument was installed on a pedestal at about 3~m above the ground
level, close to a seeing monitor and an automated weather station 
(Fig.~3). The radiometer power supply, interface unit and the control 
computer are located inside the building that houses the seeing monitor.
The system was switched on for continuous measurements of the zenith 
optical depth on 23 December 1999.

Till May 2000, the system was in effect sampling the western sky as it 
was oriented to take scans in the west to east direction. After that, it
was re-oriented to sample the northern sky with the scans being taken in
the north to south direction. This change was carried out to eliminate 
direct heating of components inside the radome by the Sun at low 
elevations and a suitable cover was fixed to provided shade in order to
reduce heating by the Sun at high elevations in the new orientation.

\section*{\underbar{System sensitivity}}

The system sensitivity expressed as the minimum detectable signal power 
($1\sigma$) is given by the following expression.
\begin{equation}
T_{\rm {rms}} = T_{\rm {sys}} / [BW \times {\rm {Integration~time}}]^{1/2} 
\end{equation}
In this case, the system temperature and receiver temperature are about
the same i.e., $T_{\rm{sys}} = T_{\rm{rx}}$ as $T_{\rm{rx}} = 13,000$~K.
\vskip 1pc
\noindent
BW is pre-detection bandwidth = $500\times 10^{6}$ Hz\\
Integration time = $90\times 10^{-3}$  s\\
System sensitivity or minimum detectable signal power $T_{rms}$ is 2~K.

\section*{\underbar{Sky scans and Analysis}}

The signal received from the sky (Fig.~4) suffers reduction in amplitude
due to absorption and there is an additive noise contribution as well. 
The basic equations relating the detected voltage and temperature are:
\begin{eqnarray}
V_{\rm{sky}}=C[T_{\rm{rx}}+T_{\rm{sky}}]\\ 
V_{\rm{ref}}=C[T_{\rm{rx}}+T_{\rm{ref}}]  
\end{eqnarray}
where, $ T_{\rm{sky}} = T_{\rm{medium}}(1-e^{-\tau_{o}\sec(z)}) 
+T_{b}e^{-\tau_{o}\sec(z)}$, $C$ is a proportionality constant or 
conversion factor, $T_{\rm{rx}}$ is receiver noise temperature and 
$T_{b}$ is background source temperature which is of the order of a few 
Kelvin.  With the assumptions that the system gain is constant or any
variation is corrected, system sensitivity is the same at all zenith 
angles, sky is stable during the scans, no background source in the beam
and $T_{\rm{medium}}=T_{\rm{ambient}}=T_{\rm{ref}}$, the above equations
can be reduced to a single linear equation and the zenith optical depth 
can be determined by a least square fit routine using the $\sec(z)$ 
dependency of the sky emission. Data up to an airmass of about 2.5 are 
used in the fit to derive the zenith opacity. Data from a higher airmass
may contain noise picked up from the ground and will also introduce 
large errors in the fitted zenith optical depth due to uncertainties in
identifying the actual zenith position. 

If the IF box temperature ($T_{\rm{heater}}$) is below $10^{\circ}$ C, 
then the scans are taken with the gain increasing linearly.  Otherwise, 
the gain variations are very slow and proportional to the slow diurnal 
variations of ambient temperature. Averaging the forward and reverse 
scans reduces the errors on the deduced optical depth due to small gain 
variations between the scans, and if the gain change is linear with time
then averaging completely cancels the gain variations. 

\begin{figure} 
\vspace{1pc}
\centerline{\includegraphics[width=24pc]{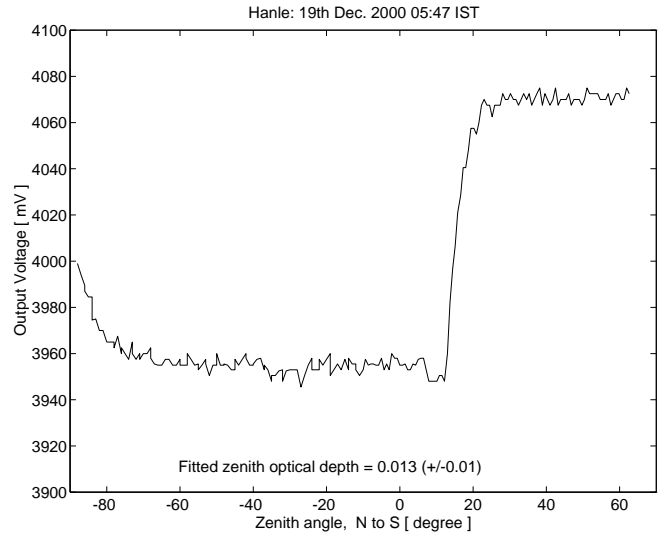}} 
\vspace{1pc}
\caption[]{A typical averaged scan, detected voltage as a function of 
zenith angle. The trace from zenith angle between $15^{\circ}$ and 
$22^{\circ}$ is the transition voltage as the mirror begins to partially
look at sky and the absorber. The higher voltage between $22^{\circ}$ 
and $60^{\circ}$ is the reference voltage when the mirror fully looks at
the absorber.}
\label{rmscan}
\end{figure}

\section*{\underbar{Results and Interpretation}}

The opacity values between 23 December 1999 and 12 May 2000 have larger 
errors due to imperfect gain cancellation between the forward and 
reverse scan pairs in a 10 minute sample.  About 10\% of the data during
this period were rejected being totally unusable.  The system was also 
unstable, working intermittently from April to August, and fully stopped
functioning by September 2000. The troubleshooting and rectification of 
bugs and also the overhauling and improvements undertaken in October 
2000 have made the system reliable since then. The best month during the
period subsequent to October 2000 was December and the observed 
opacities for this month are shown in Fig.~5, where the opacities are 
averaged over hourly intervals (7 scans including both the end points).

\begin{figure} 
\vspace{1pc}
\centerline{\includegraphics[width=24pc]{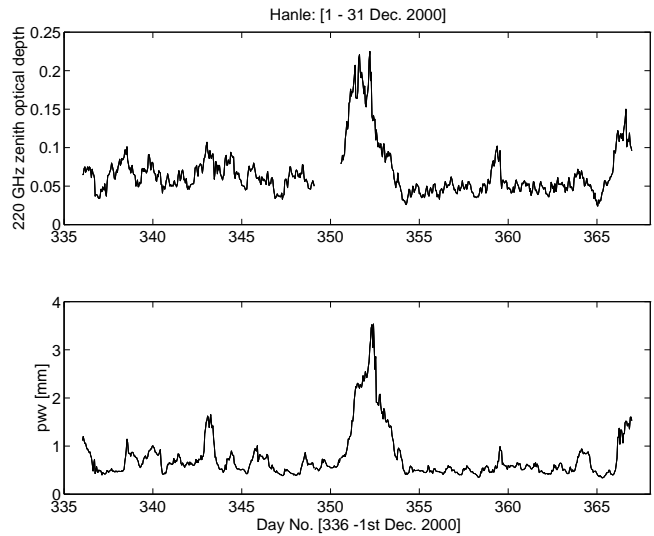}} 
\vspace{1pc}
\caption[]{The derived opacities (hourly average) during the month of 
December 2000 (top panel). The bottom panel shows the precipitable water
vapour estimated from the surface-weather-data.} 
\label{oppwv}
\end{figure}
  
The $1\sigma$ error on the 220~GHz zenith optical depth from the fit is 
of the order of 0.01 and systematic errors are expected to be under 
15\%. The systamatic errors may be contributed by the following factors.

\begin{enumerate}

\item Error in the identification of true zenith position

\item Error due to reflections between the receiver horn and membrane 
resulting in a ripple in the output voltage

\item Leftover residual non-linear gain effects 

\item The absorber temperature being higher than the temperature of the 
medium (effective atmosphere)

\end{enumerate}

The peak-to-peak errors on the determined zenith position are of the 
order $\pm 2.7^{\circ}$. 

The fitted opacity may be negative or an overestimate if the atmosphere 
was non-uniform. For example the opacities will be negative if the 
atmosphere was to be non-uniform or with clouds near zenith and may be 
an overestimate if there were clouds in the beam at low elevations.
We have neglected fitted opacity values that are negative, but retained
all the high values though some of them may be spurious. This will
lead to a slightly skewed distribution to higher opacity values, but
the median of the distribution would be affected little.

An automated weather station has been installed at Hanle in July 1996
and hourly data is available on air temperature and relative humidity
besides soil temperature, wind velocity vector, solar radiation and
rainfall.  The partial pressure of water vapour at the surface can be 
computed from the saturation water vapour pressure at the ambient 
temperature and the surface relative humidity \cite{Butler, Liebe}:
\begin{equation}
P_0=2.409\times 10^{12}RH(300/T)^4e^{-6792/T} 
\end{equation}
where, $P_0$ is the surface water vapour pressure (WVP) in microbars, 
$RH$ is the relative humidity in \%, and $T$ is the surface air 
temperature in K.  

\begin{figure} 
\vspace{1pc}
\centerline{\includegraphics[width=23pc]{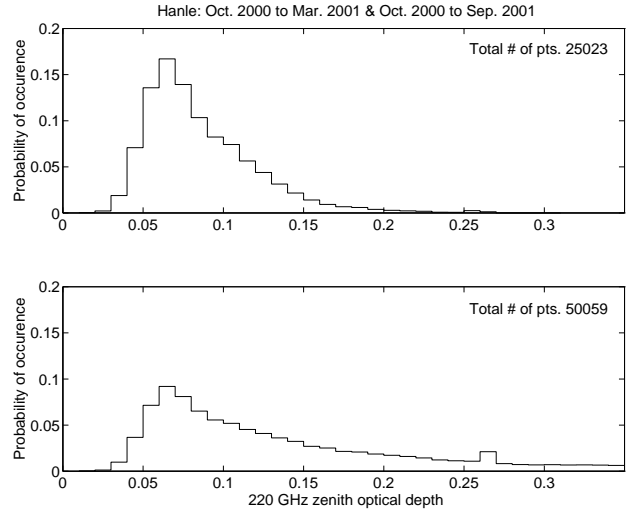}} 
\vspace{1pc}
\caption[]{Relative frequency distribution of 220~GHz zenith optical 
depth for the period 6th October 2000 to 31st March 2001 (top panel) and
6 October 2000 to 30 September 2001 (bottom panel).}
\label{opfreq}
\end{figure}

\begin{figure} 
\vspace{1pc}
\centerline{\includegraphics[width=23pc]{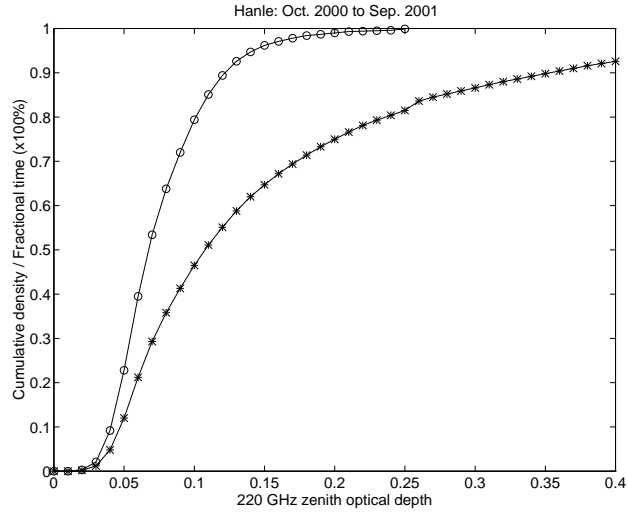}} 
\vspace{1pc}
\caption[]{Cumulative density distribution function of the 220~GHz zenith 
optical depth. $\circ$ -- distribution for the period 6 October 2000 to 
31 March 2001 i.e., winter 2000--01. * -- distribution for the period 
6 October 2000 to 30 September 2001.} 
\label{opcum}
\end{figure}

The total water vapour content (in mm) above the surface can then be 
calculated using the expression below, when the surface temperature 
lies in the 250 to 310~K range.
\begin{equation}
pwv= P_0/(3.0\times T)
\end{equation}
where, $pwv$ is the total precipitable water vapour column above the 
site in mm, and a scale height of water vapour in $H=1.5$ km is assumed.
The scale height is known to vary diurnally and seasonally, and is 
expected to be in the 1 to 2~km range. The lower panel of Fig.~5 shows 
the computed precipitable water vapour for the month of December 2000.  
A correlation between opacity and $pwv$ is clearly evident. 

The relative frequency distribution of zenith optical depth is shown in 
Fig.~6 separately for the winter months (October 2000 to March 2001), 
and the entire duration of observations (October 2000 to September 
2001). The corresponding cumulative frequency distribution functions are
shown in Fig.~7.

The 220~GHz zenith optical depth is expected to be correlated with the
precipitable water vapour above the site as was brought out by Fig.~5. 
We proceed to derive such a relationship using the (median) quartiles
of the frequency distribution of optical depth and the (median) 
quartiles of $pwv$ derived from meteorological parameters. The plot in 
Fig.~8 for the period October 2000 to September 2001 with $H$ = 1.5~km 
yields the following correlations:
\begin{eqnarray}
\tau_{220} = 0.0281+ 0.0462pwv&\\
\tau_{220} = 0.0377+ 0.0363pwv& + 0.0014pwv^2
\end{eqnarray}

\begin{figure} 
\vspace{1pc}
\centerline{\includegraphics[width=24pc]{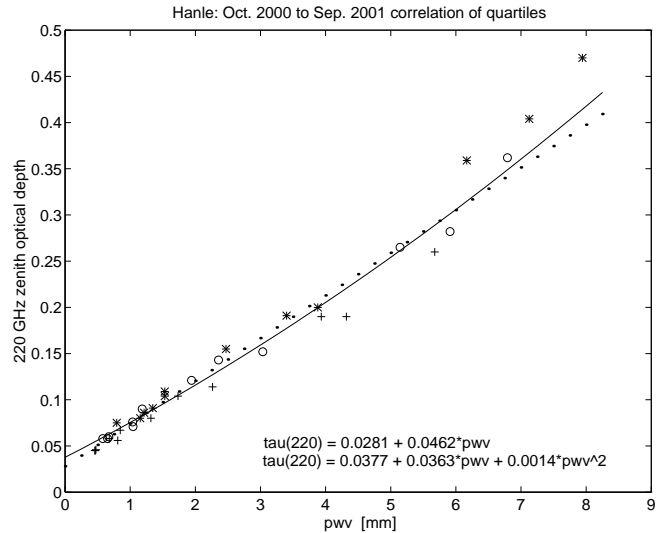}} 
\vspace{1pc}
\caption[]{A plot of the quartiles of measured opacities, and the 
quartiles of precipitable water vapour column derived from weather 
parameters for an assumed scale height of 1.5~km. $+$: first quartile;
$\circ$: median; *: third quartile. A least square linear
fit is shown as dotted line and a quadratic fit by continuous line.} 
\label{opwvq}
\end{figure}

The relationship between $pwv$ derived from 183~GHz water vapour line 
measurement (assumed $H$ = 2~km) and the 225~GHz opacity measurements at
Atacama (5000~m) is \cite{Guillermo-etal}
\begin{equation}
\tau_{225}=0.007 + 0.041pwv + 0.0009pwv^2
\end{equation}
The opacity at 220~GHz is expected to be about 19\% lower compared to 
the 225~GHz opacity (in the range 0.06 to 0.12). Thus, the relationship 
derived by us is (slightly) steeper, and has an appreciable offset. It 
is likely that the offset results from residual systematic errors 
mentioned earlier in this section.  While the slopes can be affected by 
systematic percentile errors in the opacity estimates, the major source 
of error could also be contributed by the uncertainty in the mean scale 
height. 

The early determinations of optical depth (1999 December -- 2000 May),
though having larger errors, agree with the weather derived $pwv$ in 
their quartiles. We thus believe that the early measurements are on the 
average consistent with the current set of improved data. The zenith
optical depth quartiles for both these periods are shown in Fig.~9, as 
also listed in Table~1.

\begin{table} 
\begin{tabular}{lrrrr}                                        
\hline
Month & Year & Q1 & Q2 & Q3\\
\hline\\
\\
January &2000&  0.055& 0.079& 0.105\\
	&2001&  0.045& 0.058& 0.080\\
February&2000&  0.055& 0.076& 0.098\\
	&2001&  0.045& 0.060& 0.086\\
March   &2000&  0.059& 0.082& 0.110\\
	&2001&  0.058& 0.076& 0.104\\
April   &2000&  0.074& 0.108& 0.148\\
	&2001&  0.080& 0.121& 0.155\\
May     &2000&  0.096& 0.136& 0.194\\
	&2001&  0.104& 0.143& 0.191\\
June    &2001&  0.190& 0.265& 0.359\\
July	&2001&  0.260& 0.362& 0.470\\
August  &2001&  0.190& 0.282& 0.404\\
September &2001&0.114& 0.152& 0.200\\
October &2000&  0.056& 0.071& 0.091\\
November&2000&  0.067& 0.090& 0.109\\
December&1999&   0.040& 0.060& 0.085\\
	&2000&  0.046& 0.058& 0.075\\
\hline
\end{tabular}
\caption[]{Percentile opacities at 220~GHz, IAO, Hanle}
\end{table}

\begin{figure} 
\vspace{1pc}
\centerline{\includegraphics[width=24pc]{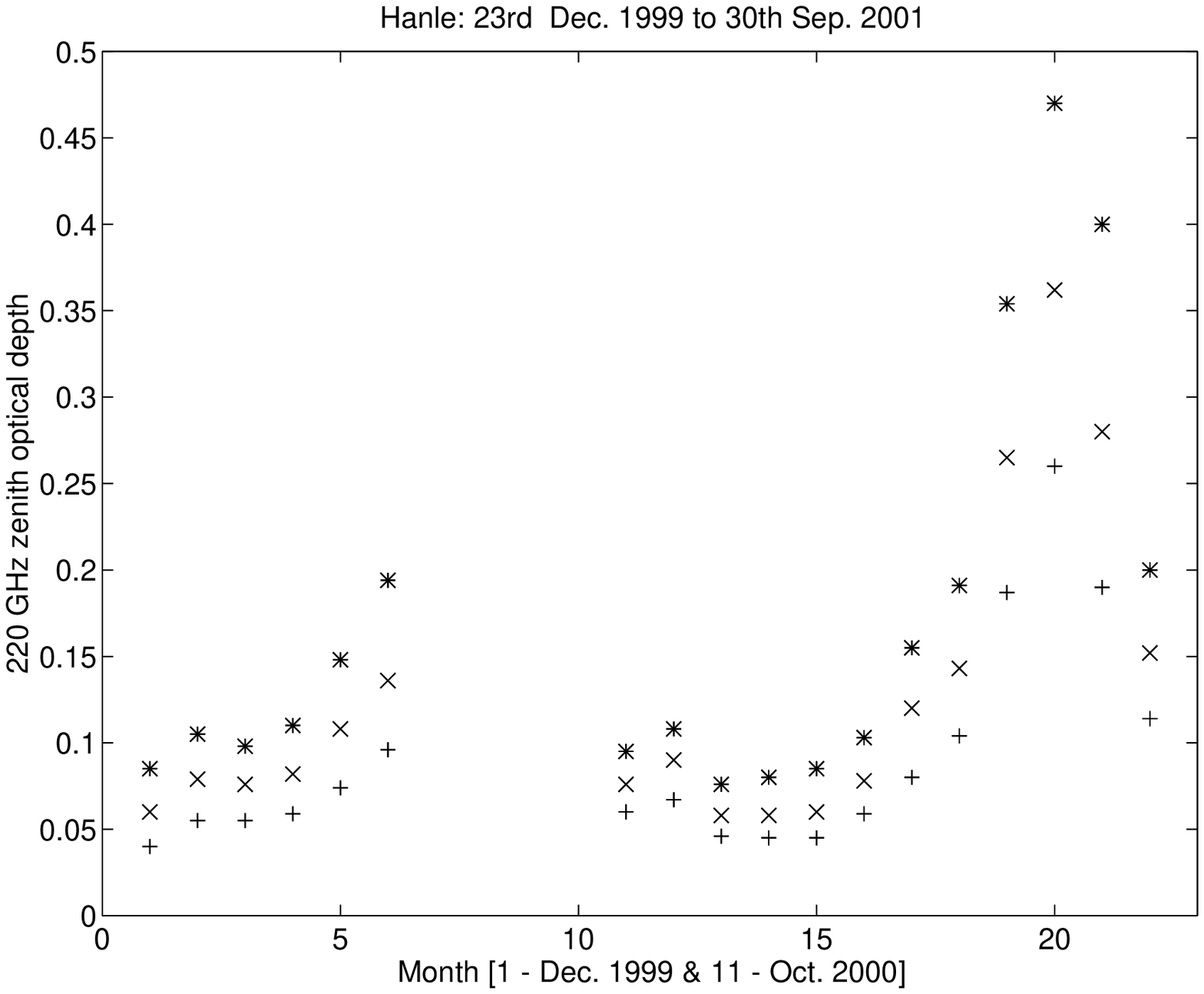}} 
\vspace{1pc}
\caption[]{Monthly quartiles of 220~GHz zenith optical depth for 
the period 23 December 1999 to 12 May 2000 and 6 October 2000 to 30 
September 2001 $+$ -- first quartile, $\times$ -- second quartile 
(median) and * -- third quartile.}     
\label{opquart}
\end{figure}

The 225~GHz zenith optical depth quartiles at Mauna Kea for the nearly
three year period (January 1997 -- October 2000) are 0.058, 0.091, 
0.153.  The opacity at at 220~GHz is expected to be lower by about 19\% 
compared to the 225~GHz opacity (in the range 0.06 to 0.12). Thus, the 
transparency of Hanle during October 2000 -- September 2001 (one year 
period: 0.065, 0.108, 0.200) is slightly higher than the annual average 
at Mauna Kea.  In Fig.~10 the one to one comparison of 225~GHz median 
opacities at Hanle and Mauna Kea for the period October 2000 to March 
2001 indicates Hanle to be slightly better than Mauna Kea in deep winter
(December -- February) and comparable over the period October -- March. 
The Mauna Kea opacities are shown in the figure after conversion to 
equivalent 220~GHz opacities using expressions from 
\cite{Holdaway-etal}. 

\begin{figure} 
\vspace{1pc}
\centerline{\includegraphics[width=24pc]{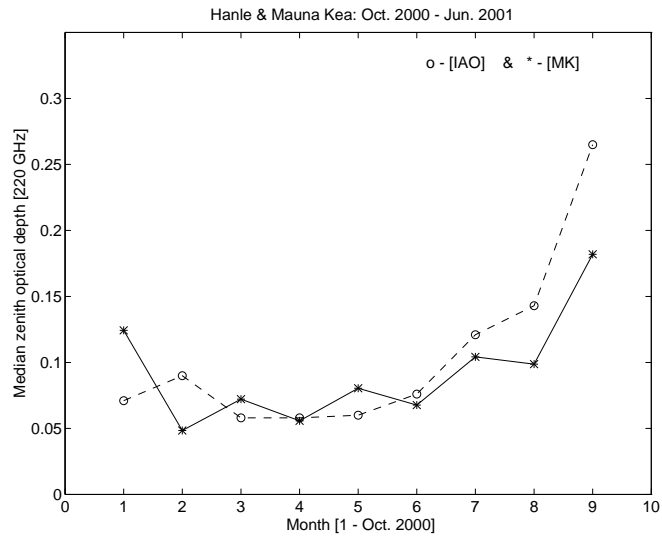}} 
\vspace{1pc}
\caption[]{Monthly  median opacities at IAO, Hanle and at Mauna Kea for 
the period October 2000 to June 2001. The Mauna Kea opacities are 
scaled to 220~GHz.} 
\label{opmed}
\end{figure}

The relation between 220~GHz opacity ($\tau_{220}$) and the 
submillimeter-wave opacity has been measured at Atacama 
\cite{Matsushita-etal}:  
\begin{eqnarray}
\tau_{492} = 21.7\times \tau_{220} + 0.270 \\
\tau_{675} = 20.7\times \tau_{220} + 0.063  
\end{eqnarray}
where, $\tau_{492}$ and $\tau_{675}$ denote the opacities at 492~GHz and
675~GHz respectively. Therefore, the 220~GHz opacity of 0.06 corresponds
to the 492~GHz opacity of 1.6, which is the actual upper limit for 
astronomical observations.

In Fig.~11 we notice that the fractional time for $\tau_{220}$ below 
0.06 at Hanle is comparable to Mauna Kea during the six winter months 
and is about 40\%. The fractional time for $\tau_{220}$ below 0.10 
(80\%) is better than at Mauna Kea (64\%) for the same period.  The 
corresponding fractional time at Mt.\ Fuji are about 45\% and about 65\%
respectively, on the average for five winter months during 1994--95.  
The numbers for opacities below 0.06 for Atacama and the South Pole are 
about 60\% and 85\% respectively, which are better than for Hanle in 
2001--02. All the numbers mentioned here are normalised to 220~GHz 
opacities, as the reported measurements were carried out at 225~GHz at 
Mauna Kea, Atacama and the South Pole.

\begin{figure} 
\vspace{1pc}
\centerline{\includegraphics[width=24pc]{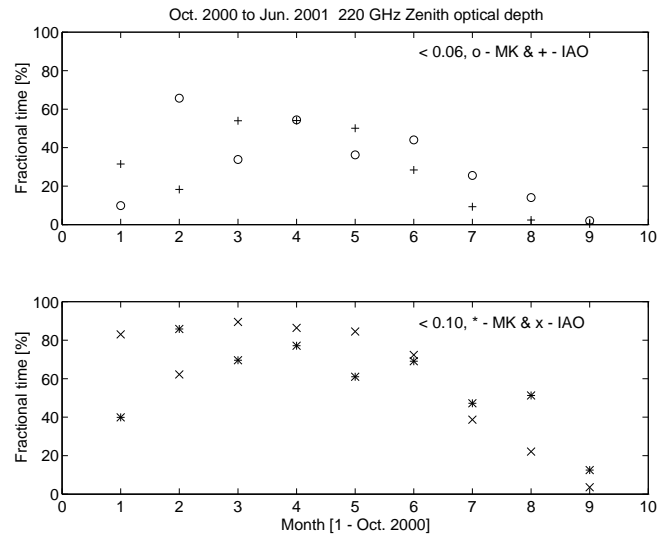}} 
\vspace{1pc}
\caption[]{Fractional time at Mauna Kea and Hanle for 220~GHz opacity 
below 0.06 and 0.10. The Mauna Kea data is corrected to be equivalent to
220~GHz opacity by appropriately scaling it.}
\label{opfrac}
\end{figure}

\begin{figure} 
\vspace{1pc}
\centerline{\includegraphics[width=24pc]{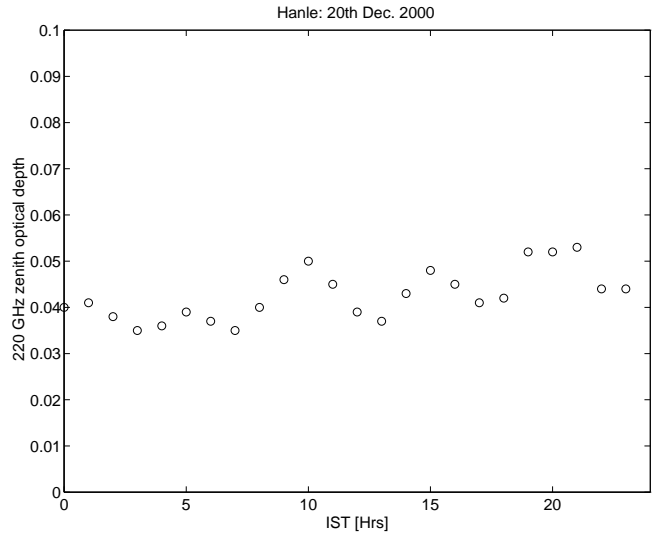}} 
\vspace{1pc}
\caption[]{Typical diurnal zenith opacity (averaged hourly) for one of 
the best days.}
\label{opday}
\end{figure}

Diurnal variations of opacity are also of interest in order to estimate 
the longest stretches of duration over which observations can be 
undertaken.  Fig.~12 shows a plot of hourly variation of opacity on 20 
December 2000. The change of opacity is rather slow and does not show 
a clear diurnal pattern. Thus it should be possible to observe 
continuously (day and night) from Hanle in the mm and sub-mm wavelength 
bands.

\section*{\underbar{Conclusions}}

It is evident from the data presented in this paper, that Hanle has
considerable advantage for observations in mm and sub-mm region. The
results presented here indicate a positive offset between measured
and expected opacities which need to be resolved in future
investigations. One needs to reduce the systematic errors in the
radiometer itself, and also to cross-calibrate the weather station data
to see if the initial calibration has drifted. It will also be useful to
launch a few radiosonde balloons at the site to estimate the detailed
structure of water vapour distribution and its scale height. 

Mt.\ Saraswati peak is in the middle of Nilamkhul plain (4250 m 
altitude) through which the Hanle river and a few underground streams 
meander. Though the amount of water flowing through is small and not 
turbulent, there is a possibility that these water sources add to the
local water vapour in the summer months. There are other peaks to the 
northeast which are possibly less affected.  A well established 
correlation between 220~GHz zenith optical depth and $pwv$ derived 
from surface relative humidity and air temperature measurements will 
help in identifying drier sites, by just setting up another weather 
station. 

We conclude that we have found and evaluated quantitatively a new site
for sub-mm astronomy in the northern hemisphere. Higher and drier sites
are available in the vicinity of Hanle for further characterisation.

\acknowledgements

A collaborative programme with remote operation of this kind could not 
be carried out successfully without active contributions from many. The 
actual list is quite long to be fully mentioned here. We express our 
appreciation of help from the following colleagues. T.\ Ito (UoT) for 
system integration and initial tests at UoT during Feb. 1999;  B.R.\ 
Madhava Rao, R.R.\ Reddy (IIA) for fabrication and erection of pedestal 
and installation of radiometer;  Angchuk Dorje, Mohinder Pal Singh and 
Tsewang Punchok (IIA) for system installation, data retrieval and 
continuous up-keep over nearly two years; 
K.B.\ Raghavendra Rao (RRI) for system integration and testing at RRI, 
before the system was transported to the Hanle site; T.K.\ Sridharan 
(CFA) for pointing out the expected low median opacities by reducing the
weather data, comparing other sites; N.\ Kumar, D.K.\ Ravindra (RRI) and
R.\ Cowsik (IIA) for continued support and encouragement. PGA thanks 
V.\ Radhakrishnan, N.V.G.\ Sarma (RRI) , Rajaram Nityananda (NCRA), 
T.K.\ Sridharan for very useful discussions and A.A.\ Deshpande for 
critical reading of the manuscript. We thank Richard A.\ Chamberlin of 
CSO for readily supporting the comparison with current 225~GHz opacity 
data at Mauna Kea.

\end{article}
\end{document}